\def \be {\begin{equation}}
\def \ee {\end{equation}}
\def \bea {\begin{eqnarray}}
\def \eea {\end{eqnarray}}
\def \bse {\begin{subequations}}
\def \ese {\end{subequations}}
\def \bde {\begin{description}}
\def \ede {\end{description}}

\def \ra {\rangle}
\def \del {\partial}

\def \dels {\partial\kern-.5em / \kern.5em}
\def \As {{A\kern-.5em / \kern.5em}}
\def \Ds {D\kern-.7em / \kern.5em}

\def \e {{\rm e}}

\def \II {I\hspace{-.1em}I\hspace{.5em}}

\def \a {\alpha}
\def \b {\beta}
\def \g {\gamma}

\def \d {\delta}
\def \eps {\epsilon}
\def \m {\mu}
\def \n {\nu}

\def \lam {\lambda}
\def \Lam {\Lambda}
\def \s {\sigma}
\def \r {\rho}

\def \th {\theta}

\def \t {\tau}

\documentclass[preprint,aps,floats]{revtex4}
\usepackage{amsmath}

\begin{document}
\title{Decoupling of Degenerate Positive-norm States in Witten's String Field Theory}
\author{Hsien-Chung Kao$^{1,2}$\footnote{hckao@phy.ntnu.edu.tw} and Jen-Chi Lee$^3$\footnote{jcclee@cc.nctu.edu.tw
}}
\address{\small $^1$Department of Physics, Tamkang University, Tamsui, Taiwan, R.O.C.}
\address{\small $^2$Department of Physics, National Taiwan Normal University, Taipei, Taiwan, R.O.C.}
\address{\small $^3$Department of Electrophysics, National Chiao Tung University, Hsinchu, Taiwan, R.O.C.}

\date{\today}

\begin{abstract}

We show that the degenerate positive-norm physical propagating
fields of the open bosonic string can be gauged to the higher rank
fields at the same mass level. As a result, their scattering
amplitudes can be determined from those of the higher spin fields.
This phenomenon arises from the existence of two types of
zero-norm states with the same Young representations as those of
the degenerate positive-norm states in the old covariant first
quantized (OCFQ) spectrum. This is demonstrated by using the
lowest order gauge transformation of Witten's string field theory
(WSFT) up to the fourth massive level (spin-five), and is found to
be consistent with conformal field theory calculation based on the
first quantized generalized sigma-model approach. In particular,
on-shell conditions of zero-norm states in OCFQ stringy gauge
transformation are found to correspond, in a one-to-one manner, to
the background ghost fields in off-shell gauge transformation of
WSFT. The implication of decoupling of scalar modes on Sen's
conjectures was also briefly discussed.

\end{abstract}
\maketitle

\setcounter{footnote}{0}

\section{Introduction}

It was pointed out more than ten years ago by Gross \cite{Gross1} that, in addition to the strong coupling regime, the most important nonperturbative regime of string theory is the high-energy stringy ($\a' \rightarrow \infty $) behavior of the theory. It is in this regime that the theory becomes very different from point particle field theory. Among many interesting stringy
 behaviors, it was believed that an infinite broken gauge symmetry get restored at energy much higher than the Planck energy. Moreover, this symmetry is powerful enough to link different string scattering amplitudes and, in principle, can be used to express all string amplitudes in terms of say dilaton amplitude.

Instead of studying stringy scattering amplitudes \cite{Gross2},
one alternative to explicitly derive stringy symmetry is to use
the generalized worldsheet sigma-model approach. In this approach,
one uses conformal field theory to calculate the equations of
motion for massive string background fields in the lowest order
weak field approximation but valid to all orders in $\a'$. Weak
field approximation is thus the appropriate approximation scheme
to study high energy symmetry of the string. An infinite set of
on-shell stringy gauge symmetry is then derived by requiring the
decoupling of both types of zero-norm physical states in the OCFQ
spectrum \cite{jclee1}. In particular, all physical propagating
states at each fixed mass level are found to form a large gauge
multiplet. This begins to show up at the second massive level
(spin-three).  Moreover, it was remarkable to discover that
\cite{jclee2}, the degenerate positive-norm physical propagating
fields of the third massive level of the open bosonic string can be gauged to the higher rank fields by the existence of zero-norm states with the same Young
representations. It was also shown \cite{jclee3} that the
scattering amplitudes of these degenerate positive-norm states can
be expressed in terms of those of higher spin states at the same
mass level through massive Ward identities. The subtlety of the scalar state pointed out in [5] will be resolved in the end of section II. This phenomenon begins to show up at the third massive level (spin-four), and was argued to be a
sigma-model n+1 loop result for the n-th massive level. These
stringy phenomena seem to be closely related to the results in
Ref.~\cite{Gross1}. In fact, an infinite number of linear
relations between the string tree-level scattering amplitudes of
different string states, similar to those claimed in
Ref.~\cite{Gross1}, can be derived by making use of an infinite number of
zero-norm states \cite{jclee3}. To claim that the decoupling
phenomenon persist for general higher levels, it would be very
important a prior to see whether one can rederive it from the
second quantized off-shell WSFT \cite{Witten1}.

Recently there is a revived interest in WSFT, mainly due to Sen's
conjecture in tachyon condensation on D-brane \cite{Sen}. It
becomes more and more clear that a second quantized field theory
of string is unavoidable especially when one wants to study higher
string modes. Thus, a cross check by both first and second
quantized approaches of any reliable string theory result would be
of great importance.  Unfortunately most of the recent researches
on string field theory were confined to the scalar modes on
identification of nonperturbative string vacuum \cite{Kostelecky}.
Our aim in this paper is to consider the gauge transformation of
all string modes with any spin and in arbitrary gauge
\cite{Taylor}. We will first prove the decoupling phenomenon of
the third massive level of open bosonic string claimed in
Ref.~\cite{jclee2} by WSFT. The result is then generalized to the
fourth massive level by both first and second quantized
approaches. This paper is organized as follow. In section II we
first summarize the previous results obtained in first quantized
approach. In section III we explicitly calculate the lowest order
gauge transformation level by level up to the third massive level
in WSFT, and compare them with those of the first quantized
approach. Some important observations will be made for ghost
fields in WSFT and zero-norm states in OCFQ spectrum. The
transformation will be separated into the matter and ghost fields
parts in WSFT. The matter part is found to be consistent with
previous calculation \cite{jclee3} based on the old covariant
string field gauge transformation of Banks and Peskin
\cite{Banks}. The ghost part is argued to correspond to the
lifting of on-shell (including on-mass-shell, gauge and traceless)
conditions of zero-norm states in the OCFQ calculation.  Section
IV is devoted to the fourth massive level. Both first and second
quantized calculations there are new and will be presented. A
brief conclusion is made in section V. The lengthy gauge
transformation of ghost fields for level four will be collected in
the appendix.

\section{Old covariant first quantized approach }

The old covariant quantization is one of the three standard
quantization schemes of string.  In addition to the physical
positive-norm propagating modes, there exist two types of physical
zero-norm states in the bosonic open string spectrum \cite{GSW}.
They are \bea &\;&{\rm Type}\; {\rm I:} \quad L_{-1}|\chi\ra,
\quad {\rm where}\; L_m |\chi\ra =0, m\ge 1, L_0|\chi\ra = 0;
\label{Cond_1}\\ &\;& {\rm Type}\; {\rm \II\hspace{-0.6em}:} \quad
\left(L_{-2}+\frac{3}{2}L^2_{-1}\right)|\tilde{\chi}\ra, \quad
{\rm where}\; L_m |\tilde{\chi}\ra =0, m\ge 1,
(L_0+1)|\tilde{\chi}\ra = 0. \label{Cond_2} \eea While type I
states have zero-norm at any spacetime dimension, type \II states
have zero-norm {\it only} at $D=26$. Their existence turns out to
be important in the following discussion.  The explicit forms of
these zero-norm states have been calculated and their Young
tabulation, together with positive-norm states, up to the third
massive level are listed in the following table.  Note that
zero-norm states are not included in the light-cone quantization.
\vskip 2cm
\begin{tabular}{|c|c|c|}
\hline
mass level&positive-norm states&zero-norm states \\
\hline
$m^2 =-2$&$\bullet$& \\
$m^2 =0$&\raisebox{0.04in}{\fbox}& $\bullet$({\rm singlet})\\
$m^2 =2$&\raisebox{0.04in}{\fbox}\hspace{-0.005in}\raisebox{0.04in}{\fbox}& \raisebox{0.04in}{\fbox},$\bullet$\\
$m^2 =4$&$\raisebox{0.04in}{\fbox}\hspace{-0.005in}\raisebox{0.04in}{\fbox}\hspace{-0.006in}\raisebox{0.04in}{\fbox},\raisebox{0.06in}{\fbox}\hspace{-0.094in}\raisebox{-.029in}{\fbox}$&$\raisebox{0.04in}{\fbox}\hspace{-0.005in}\raisebox{0.04in}{\fbox}, 2\times\raisebox{0.04in}{\fbox},\bullet$ \\
$m^2 =6$&$\raisebox{0.04in}{\fbox}\hspace{-0.005in}\raisebox{0.04in}{\fbox}\hspace{-0.005in}\raisebox{0.04in}{\fbox}\hspace{-0.005in}\raisebox{0.04in}{\fbox},\raisebox{0.06in}{\fbox}\hspace{-0.094in}\raisebox{-.029in}{\fbox}\hspace{-0.006in}\raisebox{0.06in}{\fbox},\raisebox{0.04in}{\fbox}\hspace{-0.005in}\raisebox{0.04in}{\fbox},\bullet$& $\raisebox{0.04in}{\fbox}\hspace{-0.006in}\raisebox{0.04in}{\fbox}\hspace{-0.006in}\raisebox{0.04in}{\fbox},\raisebox{0.06in}{\fbox}\hspace{-0.094in}\raisebox{-.029in}{\fbox},2\times\raisebox{0.04in}{\fbox}\hspace{-0.005in}\raisebox{0.04in}{\fbox},3\times\raisebox{0.04in}{\fbox},2\times\bullet$\\
&&\\
\hline
\end{tabular}
\vskip 10pt \vbox{\small \noindent Table.1 OCFQ spectrum of open
bosonic string.}

\vskip 1cm
It was demonstrated in the first order weak field
approximation that for each zero-norm state there corresponds an on-shell
gauge transformation for the positive-norm background
field ($\alpha'\equiv \frac{1}{2}$) \cite{jclee1}:
\bse
\label{rule_A}
\bea
&\;& \hspace{-8.2cm} m^2=0: \quad \delta A_\m = \del_\m\th; \label{rule_Aa} \\
&\;& \hspace{-6.15cm} \del^2\th =0. \label{rule_Ab}
\eea
\ese
\vspace{-1.1cm}
\bse
\label{rule_B1}
\bea
&\;& \hspace{-5.9cm} m^2=2: \quad \delta B_{\m\nu} = \del_{(\m}\th_{\n)}; \\
&\;& \hspace{-3.85cm} \del^\m\th_\m =0, (\del^2-2)\th_\m=0. \label{rule_B1b}
\eea
\ese
\vspace{-1.1cm}
\bse
\label{rule_B2}
\bea
&\;& \hspace{-3.85cm} \delta B_{\m\nu} = \frac{3}{2}\del_{\m}\del_\n\th - \frac{1}{2}\eta_{\m\nu}\th; \\
&\;& \hspace{-3.85cm} (\del^2-2)\th=0. \label{rule_B2b}
\eea
\ese
\vspace{-1.1cm}
\bse
\label{rule_C1}
\bea
&\;& \hspace{-4.5cm} m^2=4: \quad \delta C_{\m\nu\lam} = \del_{(\m}\th_{\n\lam)};\\
&\;& \hspace{-2.45cm} \del^\m\th_{\m\nu} = \th^{\;\m}_\m = 0,\; (\del^2-4)\th_{\m\nu} =0.  \label{rule_C1b}
\eea
\ese
\vspace{-1.1cm}
\bse
\label{rule_C2}
\bea
&\;& \hspace{-3.05cm} \delta C_{\m\nu\lam} =  \frac{5}{2}\del_{(\m}\del_\n\th^1_{\lam)}-\eta_{(\m\nu}\th^1_{\lam)}; \\
&\;& \hspace{-3.05cm} \del^\m\th^1_{\m} = 0,\; (\del^2-4)\th^1_{\m} =0. \label{rule_C2b}
\eea
\ese
\vspace{-1.1cm}
\bse
\label{rule_C3}
\bea
&\;& \delta C_{\m\nu\lam} = \frac{1}{2}\del_{(\m}\del_\n\th^2_{\lam)}-2\eta_{(\m\nu}\th^2_{\lam)},\; \delta C_{[\m\nu]} = 9 \del_{[\m}\th^2_{\n]}; \\
&\;& \del^\m\th^2_{\m} = 0,\; (\del^2-4)\th^2_{\m} =0. \label{rule_C3b}
\eea
\ese
\vspace{-1.1cm}
\bse
\label{rule_C4}
\bea
&\;& \hspace{-2.65cm} \delta C_{\m\nu\lam} = \frac{3}{5}\del_{\m}\del_\n\del_\lam\th -
\frac{1}{5}\eta_{(\m\nu}\del_{\lam)}\th; \\
&\;& \hspace{-2.65cm} (\del^2-4)\th =0. \label{rule_C4b}
\eea
\ese

In the above equations, A, B, C are positive-norm background
fields, $\th$'s represent zero-norm background fields, and
$\del^2\equiv \del^\m \del_\m$. There are on-mass-shell, gauge and
traceless conditions on the transformation parameters $\th$'s,
which will correspond to BRST ghost fields in a one-to-one manner
in WSFT as will be discussed in the next section.
Eq~(\ref{rule_A}) is of course the usual on-shell gauge
transformation, and eq~(\ref{rule_B2}) is the first residual
stringy gauge symmetry. Note that $\th^1_\m$ and $\th^2_\m$ in
eqs~(\ref{rule_C2}) and (\ref{rule_C3}) are some linear
combination of the original type I and type \II vector zero-norm
states calculated by eqs~(\ref{Cond_1}) and (\ref{Cond_2}). It is
interesting to see that eq~(\ref{rule_C3}) implies that the two
second massive level modes $C_{\m\nu\lam}$ and $C_{[\m\nu]}$ form
a larger gauge multiplet \cite{jclee1}. This is a generic feature
for higher massive level and had also been justified from S-matrix
point of view \cite{jclee4}.  One might want to generalize the
calculation to the second order weak field to see the inter-mass
level symmetry.  This however suffers from the so-called
non-perturbative non-renormalizability of 2-d $\sigma$-model and
one is forced to introduce infinite number of counter-terms to
preserve the worldsheet conformal invariance \cite{Das}.

Instead of calculating the stringy gauge symmetry at level $m^2=6$, we will only concentrate on the equation of motion.  It was discovered that an even more interesting phenomenon begins to show up at this mass level.  Take the energy-momentum tensor on the worldsheet boundary in the first order weak field approximation to be of the following form.

\bea T(\t) = &\;& -\frac{1}{2}\eta_{\m\nu}\del_\t X^\m \del_\t
X^\n + D_{\m\nu\a\b} \del_\t X^\m \del_\t X^\n \del_\t X^\a
\del_\t X^\b + D_{\m\nu\a} \del_\t X^\m \del_\t X^\n \del_\t^2
X^\a \nonumber \label{T++}\\ &\;& + D^0_{\m\nu} \del_\t^2 X^\m
\del_\t^2 X^\n + D^1_{\m\nu} \del_\t X^\m \del_\t^3 X^\n + D_{\m}
\del_\t^4 X^\m, \eea where $\t$ is worldsheet time, $X \equiv
X(\t).$  This is the  most general worldsheet coupling in the
generalized $\s$-model approach consistent with vertex operator
consideration \cite{Weinberg}.  The conditions to cancel all
q-number worldsheet conformal anomalous terms correspond to
cancelling all kinds of loop divergences \cite{Labastida} up to
the four loop order in the 2-d conformal field theory.  It is
easier to use $T\cdot T$ operator-product calculation and the
conditions read \cite{jclee2} 
\bse 
\bea 
&\;& 2\del^\m D_{\m\nu\a\b} - D_{(\n\a\b)} = 0, \label{Eq_D1}\\ &\;& \del^\m
D_{\m\nu\a} - 2D^0_{\n\a} - 3D^1_{\n\a} = 0, \label{Eq_D2}\\ &\;&
\del^\m D^1_{\m\nu} - 12D_{\n} = 0, \label{Eq_D3}\\ &\;&
3D^\m_{\;\;\m\nu\a} + \del^\m D_{\nu\a\m} - 3D^1_{(\n\a)} = 0,
\label{Eq_D4}\\ &\;& D^\m_{\;\;\m\nu} + 4\del^\m D^0_{\m\nu} -
24D_{\n} = 0, \label{Eq_D5}\\ &\;& 2D_{\m\nu}^{\quad\nu} +
3\del^\n D^1_{\m\nu} - 12D_{\m} = 0, \label{Eq_D6}\\ &\;&
2D^{0\;\m}_{\m} + 3D^{1\;\m}_{\m} + 12\del^\m D_{\m} = 0,
\label{Eq_D7}\\ &\;& (\del^2 - 6) \phi = 0.\label{Eq_D8} 
\eea 
\ese
Here, $\phi$ represents all background fields introduced in
eq~(\ref{T++}).  It is now clear through (\ref{Eq_D2}) and
(\ref{Eq_D4}) that both $D^0_{\m\n}$ and $D^1_{(\m\n)}$ can be
expressed in terms of $D_{\m\n\a\b}$ and $D_{\m\n\a}$.
$D^1_{[\m\n]}$ can be expressed in terms of $D_{\m\n\a\b}$ and
$D_{\m\n\a}$ by (\ref{Eq_D2}). Equations (\ref{Eq_D1}) and
(\ref{Eq_D3}) imply that $D_{(\m\n\a)}$ and $D_\m$ can also be
expressed in terms of $D_{\m\n\a\b}$ and mixed-symmetric
$D_{\m\n\a}$.  Finally eqs~(\ref{Eq_D5})-(\ref{Eq_D7}) are the
gauge conditions for $D_{\m\n\a\b}$ and mixed-symmetric
$D_{\m\n\a}$ after substituting $D^0_{\m\n}, D^1_{\m\n}$ and
$D_\m$ in terms of $D_{\m\n\a\b}$ and mixed symmetric
$D_{\m\n\a}$.  The remaining scalar particle has automatically
been gauged to higher rank fields since eq~(\ref{T++}) is already
the most general form of background-field coupling.  This means
that the degenerate spin two and scalar positive-norm states can
be gauged to the higher rank fields $D_{\m\n\a\b}$ and
mixed-symmetric $D_{\m\n\a}$ in the first order weak field
approximation. In fact, for instance, it can be explicitly shown
\cite{jclee3} that the scattering amplitude involving the
positive-norm spin-two state can be expressed in terms of those of
spin-four and mixed-symmetric spin-three states due to the
existence of a type I and a type II spin-two zero-norm states. The
subtlety of the scalar state scattering amplitude pointed out in
[5] can be resolved in the following way. Take a representative of the
scalar state to be [14] 
\bea 
&\;& :\left\{ -(\eta_{\m\n}+
\frac{13}{3} k_\m k_\n)\del_z^2 X^\m \del_z^2 X^\n -i(\frac{20}{9}
k_\m k_\n k_\r +\frac{2}{3} k_\m \eta_{\n\r} +\frac{13}{3}k_\r
\eta_{\m\n})\del_z X^\m \del_z X^\n\del_z^2 X^\r \right.\nonumber
\\ &\;& \quad\; \left.+ (\frac{23}{81} k_\m k_\n k_\r k_\s
+\frac{32}{27} k_\m k_\n \eta_{\r\s} +\frac{19}{18}\eta_{\m\n}
\eta_{\r\s})\del_z X^\m \del_z X^\n\del_z X^\r \del_z X^\s
\right\} \e^{ik X(z)}:. \nonumber 
\eea 
It turns out that one cannot gauge away the first term in the above 
equation by using the two scalar zero-norm states. However we already known 
the amplitude corresponding to $\del_z^2 X^\m \del_z^2 X^\n$ are fixed by
those of the spin-four and mixed-symmetric spin-three state. The totally
symmetric spin-three amplitude corresponding to the totally
symmetric spin-three part of the second term, $\del_z X^{(\m} \del_z
X^\n\del_z^2 X^{\r)}$, can be fixed by the spin-four amplitude due to
the existence of the totally symmetric spin-three zero-norm state.
As a result, the scalar state scattering amplitude is again
fixed by the amplitudes of spin-four and mixed-symmetric
spin-three states. Although all the four-point amplitudes considered in Ref. [5]
contain three tachyons, the argument can be easily
generalized to more general amplitudes. This is very different
from the analysis of lower massive levels where all positive-norm
states have independent scattering amplitudes. Presumably, this
decoupling phenomenon comes from the ambiguity in defining
positive-norm states due to the existence of zero-norm states in
the same Young representations. We will justify this decoupling by
WSFT in the next section. Finally one expects this decoupling to
persist even if one includes the higher order corrections in weak
field approximation, as there will be even stronger relations
between background fields order by order through iteration.

\section{Witten string field theory approach}

It would be much more convincing if one can rederive the stringy
phenomena discussed in the previous section from WSFT.  Not only can
one compare the first quantized string with the second
quantized string, but also the old covariant quantized
string with the BRST quantized string.  Although the calculation
is lengthy, the result, as we shall see, are still controllable by
utilizing the results from first quantized approach in
section \II.  There exist important consistency checks of first
quantized string results from WSFT in the literature, e.g. the
rederivation of Veneziano and Kubo-Nielson amplitudes from WSFT
\cite{Giddings}. In some stringy cases, calculations can only
be done in string field theory approach. For example, the recently
developed pp-wave string amplitudes can only be calculated in the
light-cone string field theory\cite{Volovich}. Sen's recent
conjectures of tachyon condensation on D-brane again were mostly
justified by string field theory.  Therefore, a consistent check
by both first and second quantized approaches of any reliable
string results would be of great importance.

The infinitesimal gauge transformation of WSFT is
\be
\delta \Phi = Q_B \Lam + g_0(\Phi*\Lam - \Lam*\Phi). \label{d_Phi}
\ee
To compare with our first quantized results in section
\II\hspace{-0.2cm}, we only need to calculate the first term on the
right hand side of eq~(\ref{d_Phi}).  Up to the second massive
level, $\Phi$ and $\Lam$ can be expressed as
\bea
\Phi =
\biggl\{\hspace{-0.5cm} &\;& \phi(x) + i A_\m (x) \a^\m_{-1} +
\a(x) b_{-1} c_0 - B_{\m\n}(x) \a^\m_{-1}\a^\n_{-1} + i B_\m (x)
\a^\m_{-2} \nonumber\\ &\;& + i\b_\m (x) \a^\m_{-1} b_{-1} c_0 +
\b^0(x) b_{-2}c_0 + \b^1(x) b_{-1} c_{-1} \nonumber\\ &\;& - i
C_{\m\n\lam}(x) \a^\m_{-1}\a^\n_{-1}\a^\lam_{-1} - C_{\m\n}(x)
\a^\m_{-2} \a^\n_{-1} + iC_\m (x) \a^\m_{-3} \nonumber\\ &\;& -
\g_{\m\n}(x) \a^\m_{-1} \a^\n_{-1} b_{-1} c_0 + i\g^0_\m(x)
\a^\m_{-1} b_{-2} c_0 + i\g^1_\m(x) \a^\m_{-1} b_{-1} c_{-1} +
i\g^2_\m(x) \a^\m_{-2} b_{-1} c_0 \nonumber\\ &\;& + \g^0(x)
b_{-3} c_0 + \g^1(x) b_{-2} c_{-1} + \g^2(x) b_{-1} c_{-2}
\biggr\}c_1 |k\ra, \\ \Lam = \biggl\{\hspace{-0.5cm} &\;&
\eps^0(x) b_{-1} - \eps^0_{\m\n}(x) \a^\m_{-1} \a^\n_{-1} b_{-1} +
i\eps^0_{\m}(x) \a^\m_{-1} b_{-1} + i\eps^1_{\m}(x) \a^\m_{-2}
b_{-1} + i\eps^2_{\m}(x) \a^\m_{-1} b_{-2} \nonumber\\ &\;& +
\eps^1(x) b_{-2} + \eps^2(x) b_{-3} + \eps^3(x) b_{-1}b_{-2} c_0
\biggr\} |\Omega\ra.
\eea
where $\Phi$ and $\Lam$ are restricted
to ghost number 1 and 0 respectively, and the BRST charge is
\be
Q_B = \sum^\infty_{n=-\infty} L^{\rm matt}_{-n} c_n +
\sum^\infty_{m,n=-\infty}\frac{m-n}{2}:c_m c_n b_{-m-n}: -c_0.
\label{BRST_C} \ee The transformation one gets for each mass level
are \bse\bea &\;& \hspace{-5.7cm} m^2=0, \quad\: \d A_\m = \del_\m
\eps^0,\qquad\\ &\;& \hspace{-3.7cm}\d\, \a = \frac{1}{2} \del^2
\eps^0;\label{rule_AAb} \eea\ese \vspace{-1.0cm} \bse\bea &\;&
\hspace{-4.1cm}  m^2=2,  \quad\: \d B_{\m\n} = -
\del_{(\m}\eps^0_{\n)} - \frac{1}{2} \eps^1\eta_{\m\n}, \\ &\;&
\hspace{-2.1cm}\d B_\m = -  \del_{\m} \eps^1 + \eps^0_\m ,\\ &\;&
\hspace{-2.1cm}\d \b_\m = \frac{1}{2}(\del^2 - 2)\eps^0_\m,
\label{rule_BBc}\\ &\;& \hspace{-2.1cm}\d \b^0 =
\frac{1}{2}(\del^2 -2)\eps^1, \label{rule_BBd}\\ &\;&
\hspace{-2.1cm}\d \b^1 = -  \del^\m \eps^0_\m - 3 \eps^1;
\label{rule_BBe} \eea\ese \vspace{-1.0cm} \bse\bea m^2=4, &\;& \d
C_{\m\n\lam} = -  \del_{(\m} \eps^0_{\n\lam)} - \frac{1}{2}
\eps^2_{(\m} \eta_{\n\lam)}, \label{rule_CCa}\\ &\;& \d C_{[\m\n]}
= -  \del_{[\n} \eps^1_{\m]} -  \del_{[\m} \eps^2_{\n]},
\label{rule_CCc}\\ &\;& \d C_{(\m\n)} = -  \del_{(\n} \eps^1_{\m)}
-  \del_{(\m} \eps^2_{\n)} + 2\eps^0_{\m\n} - \eps^2 \eta_{\m\n},
\label{rule_CCb}\\ &\;& \d C_{\m} = -  \del_{\m} \eps^2 + 2
\eps^1_{\m} + \eps^2_{\m}, \label{rule_CCd} \\ &\;& \d \g_{\m\n} =
\frac{1}{2}(\del^2 - 4) \eps^0_{\m\n} - \frac{1}{2} \eps^3
\eta_{\m\n}, \label{rule_CCe}\\ &\;& \d \g^0_\m = \frac{1}{2}(
\del^2 - 4)\eps^2_{\m} +  \del_{\m} \eps^3, \label{rule_CCf}\\
&\;& \d \g^1_\m = - 2 \del^\n \eps^0_{\n\m} - 2\eps^1_{\m} -
3\eps^2_{\m}, \label{rule_CCg}\\ &\;& \d \g^2_\m = \frac{1}{2}(
\del^2 - 4)\eps^1_{\m} -  \del_\m \eps^3, \label{rule_CCh}\\ &\;&
\d \g^0 = \frac{1}{2}( \del^2 - 4)\eps^2 - \eps^3,
\label{rule_CCi}\\ &\;& \d \g^1 = -  \del^\m \eps^2_{\m} - 4\eps^2
- 2\eps^3, \label{rule_CCj}\\ &\;& \d \g^2 = - 2 \del^\m
\eps^1_{\m}  - 5\eps^2 + 4\eps^3 + \eps^{0\;\m}_{\m}.
\label{rule_CCk}
\eea
\ese
It is interesting to note that
eq~(\ref{rule_AAb}) corresponds to the lifting of on-mass-shell
condition in eqs~(\ref{rule_Ab}). Meanwhile (\ref{rule_BBc}) and
(\ref{rule_BBd}) correspond to on-mass-shell condition in
(\ref{rule_B2b}) and (\ref{rule_B1b}); eq~(\ref{rule_BBe})
corresponds to the gauge condition in (\ref{rule_B1b}). Similar
correspondence applies to level $m^2=4$. Eqs~(\ref{rule_CCe}),
(\ref{rule_CCf}), (\ref{rule_CCh}) and (\ref{rule_CCi}) correspond
to on-mass-shell conditions in eqs~(\ref{rule_C1b}),
(\ref{rule_C2b}), (\ref{rule_C3b}) and (\ref{rule_C4b}).
Eqs~(\ref{rule_CCg}), (\ref{rule_CCj}) and (\ref{rule_CCk})
correspond to gauge conditions in eqs~(\ref{rule_C1b}),
(\ref{rule_C2b}) and (\ref{rule_C3b}). The traceless condition in
(\ref{rule_C1b}) corresponds to the trace part of
eq~(\ref{rule_CCe}). Also, only zero-norm state transformation
parameters appear on the r.h.s. of matter transformation A,B,C,
and all ghost transformations correspond, in a one-to one manner,
to the lifting of on-shell conditions (including on-mass-shell,
gauge and traceless conditions) in the OCFQ approach. These
important observations simplify the demonstration of decoupling of
degenerate positive-norm states at higher mass
levels, $m^2=6$ and $m^2=8$ more specifically, in WSFT, as will be discussed in the
rest of this paper.

For $m^2=4$, it can be checked that only $C_{\m\n\lam}$ and
$C_{[\m\n]}$ are dynamically independent and they form a gauge multiplet, which is consistent with result of first quantized calculation presented in section \II.

We now show the decoupling phenomenon for the third massive
level $m^2=6$, in which $\Phi$ and $\Lam$ can be expanded as \bea
\Phi_4 = \biggl\{\hspace{-0.5cm} &\;& D_{\m\n\a\b}(x)
\a^\m_{-1}\a^\n_{-1}\a^\a_{-1}a^\b_{-1} - iD_{\m\n\a}(x)
\a^\m_{-1}\a^\n_{-1}\a^\a_{-2} - D^0_{\m\n}(x) \a^\m_{-2}
\a^\n_{-2} - D^1_{\m\n}(x) \a^\m_{-1} \a^\n_{-3} \nonumber\\ &\;&
+ iD_\m (x) \a^\m_{-4} - i\xi_{\m\n\a}(x) \a^\m_{-1} \a^\n_{-1}
\a^\a_{-1} b_{-1} c_0 - \xi^0_{\m\n}(x) \a^\m_{-2} \a^\n_{-1}
b_{-1} c_0 - \xi^1_{\m\n}(x) \a^\m_{-1} \a^\n_{-1} b_{-2} c_0
\nonumber \\ &\;& - \xi^2_{\m\n}(x) \a^\m_{-1} \a^\n_{-1} b_{-1}
c_{-1} + i\xi^0_\m(x) \a^\m_{-3} b_{-1} c_0 + i\xi^1_\m(x)
\a^\m_{-2} b_{-2} c_0 + i\xi^2_\m(x) \a^\m_{-1} b_{-3} c_0
\nonumber \\ &\;& + i\xi^3_\m(x) \a^\m_{-2} b_{-1} c_{-1} +
i\xi^4_\m(x) \a^\m_{-1} b_{-2} c_{-1} + i\xi^5_\m(x) \a^\m_{-1}
b_{-1} c_{-2} + \xi^0(x) b_{-4} c_0 + \xi^1(x) b_{-3} c_{-1}
\nonumber\\ &\;& + \xi^2(x) b_{-2} c_{-2} + \xi^3(x) b_{-1} c_{-3}
+ \xi^4(x) b_{-2} b_{-1} c_{-1} c_{-0}  \biggr\}c_1 |k\ra, \\
\Lam_4 = \biggl\{\hspace{-0.5cm} &\;& -i\eps^0_{\m\n\a}(x)
\a^\m_{-1} \a^\n_{-1} \a^\a_{-1} b_{-1} - \eps^1_{\m\n}(x)
\a^\m_{-2} \a^\n_{-1} b_{-1} - \eps^2_{\m\n}(x) \a^\m_{-1}
\a^\n_{-1} b_{-2} + i\eps^3_{\m}(x) \a^\m_{-3} b_{-1} \nonumber\\
&\;& + i\eps^4_{\m}(x) \a^\m_{-2} b_{-2} + i\eps^5_{\m}(x)
\a^\m_{-1} b_{-3} + i\eps^6_{\m}(x) \a^\m_{-1} b_{-2} b_{-1} c_0 +
\eps^4(x) b_{-4} \nonumber\\ &\;& + \eps^5(x) b_{-3}b_{-1} c_0 +
\eps^6(x) b_{-2}b_{-1} c_{-1} \biggr\} |\Omega\ra. \eea The
transformations for the matter part are \bse \label{rule_DD} \bea
&\;& \d D_{\m\n\a\b} = -  \del_{(\b} \eps^0_{\m\n\a)} -
\frac{1}{2} \eps^2_{(\m\n} \eta_{\a\b)}, \label{rule_DDm_a} \\
&\;& \d D_{\m\n\a} = - \del_{(\m} \eps^1_{|\a|\n)} - \del_{\a}
\eps^2_{\n\m} + 3 \eps^0_{\m\n\a} - \frac{1}{2} \eps^4_{\a}
\eta_{\n\m} - \eps^5_{(\m} \eta_{\n)\a}, \label{rule_DDm_b} \\
&\;& \d D^1_{[\m\n]} = -  \del_{[\m} \eps^3_{\n]} -  \del_{[\n}
\eps^5_{\m]} + 2\eps^1_{[\n\m]}, \label{rule_DDm_c} \\ &\;& \d
D^1_{(\m\n)} = -  \del_{(\m} \eps^3_{\n)} -  \del_{(\n}
\eps^5_{\m)} + 2\eps^1_{(\n\m)} + 2\eps^2_{\m\n} - \eps^4
\eta_{\m\n}, \label{rule_DDm_d} \\ &\;& \d D^0_{\m\n} = -
\del_{(\m} \eps^4_{\n)} + \eps^1_{(\m\n)} - \frac{1}{2} \eps^4
\eta_{\m\n}, \label{rule_DDm_e} \\ &\;& \d D_{\m} = -  \del_\m
\eps^4 + 3\eps^3_\m + 2\eps^4_\m + \eps^5_{\m}. \label{rule_DDm_f}
\eea\ese
It can be checked from eqs~(\ref{rule_DD}) that only
$D_{\m\n\a\b}$ and mixed-symmetric $D_{\m\n\a}$ cannot be gauged
away, which is consistent with the result of the first quantized
approach in sec. II.  That is , the spin-two and scalar positive-norm
physical propagating modes can be gauged to $D_{\m\n\a\b}$ and mixed
symmetric $D_{\m\n\a}$. In fact, $D_{\m\n\a}, D^1_{[\m\n]}, D^1_{(\m\n)},
D^0_{\m\n}$ and $D_\m$ can be gauged away by $\eps^{0}_{\m\n\lam},
\eps^{1}_{[\m\n]}, \eps^{1}_{(\m\n)}, \eps^{2}_{\m\n}$ and one of
the vector parameters, say $\eps^{3}_{\m}$.  The rest,
$\eps^{4}_{\m}, \eps^{5}_{\m}$ and $\eps^{4}$ are gauge artifacts
of $D_{\m\n\a\b}$ and mixed symmetric $D_{\m\n\a}$.

The transformation for the ghost part are
\bse
\label{rule_DDg}
\bea
&\;& \d \xi_{\m\n\a} = \frac{1}{2}(\del^2 - 6) \eps^0_{\m\n\a} - \frac{1}{2} \eps^6_{(\m} \eta_{\n\a)}, \label{rule_DDg_a}\\
&\;& \d \xi^0_{[\m\n]} = \frac{1}{2}( \del^2 - 6)\eps^1_{[\m\n]} -  \del_{[\m} \eps^6_{\n]}, \label{rule_DDg_b} \\
&\;& \d \xi^0_{(\m\n)} = \frac{1}{2}( \del^2 - 6)\eps^1_{(\m\n)} -  \del_{(\m} \eps^6_{\n)} + \eps^5 \eta_{\m\n}, \label{rule_DDg_c} \\
&\;& \d \xi^1_{\m\n} = \frac{1}{2}( \del^2 - 6)\eps^2_{\m\n} +  \del_{(\n} \eps^6_{\m)}, \label{rule_DDg_d} \\
&\;& \d \xi^2_{\m\n} = - 3 \del^{\a} \eps^0_{\m\n\a} - 2\eps^1_{(\m\n)} - 3\eps^2_{\m\n} - \frac{1}{2} \eps^6 \eta_{\m\n}, \label{rule_DDg_e} \\
&\;& \d \xi^0_\m = \frac{1}{2}( \del^2 - 6)\eps^3_{\m} -  \del_{\m} \eps^5 + \eps^6_{\m},\\
&\;& \d \xi^1_\m = \frac{1}{2}( \del^2 - 6)\eps^4_{\m} - \eps^6_{\m},\\
&\;& \d \xi^2_\m = \frac{1}{2}( \del^2 - 6)\eps^5_{\m} +  \del_{\m} \eps^5 - \eps^6_{\m},\\
&\;& \d \xi^3_\m = -  \del^\n \eps^1_{\m\n} -  \del_\m \eps^6 - 3\eps^3_{\m} - 3\eps^4_{\m},\\
&\;& \d \xi^4_\m = 2 \del^\n \eps^2_{\m\n} +  \del_\m \eps^6 - 2\eps^4_{\m} - 4\eps^5_{\m} - 2\eps^6_{\m},\\
&\;& \d \xi^5_\m = - 2 \del^\n \eps^1_{\n\m} - 3\eps^3_{\m} - 5\eps^5_{\m} + 4\eps^6_{\m} + 3\eps^{0\;\;\n}_{\m\n}, \\
&\;& \d \xi^0 = \frac{1}{2}( \del^2 - 6)\eps^4 - 2\eps^5,\\
&\;& \d \xi^1 = -  \del^\m \eps^5_{\m} - 5\eps^4 - 2\eps^5 - \eps^6,\\
&\;& \d \xi^2 = - 2 \del^\m \eps^4_{\m} - 6\eps^4 - 3\eps^6 + \eps^{2\;\m}_{\m}, \\
&\;& \d \xi^3 = - 3 \del^\m \eps^3_{\m} - 7\eps^4 + 6\eps^5 + 5\eps^6 + 2\eps^{1\;\m}_{\m}, \\
&\;& \d \xi^4 = \frac{1}{2}( \del^2 - 6)\eps^6 +  \del^\m \eps^6_{\m} + 4\eps^5.
\eea
\ese

There are nine on-mass-shell conditions, which contains a
symmetric spin three,an antisymmetric spin two, two symmetric spin
two, three vector and two scalar fields, and seven gauge
conditions which amounts to sixteen equations in (\ref{rule_DDg}).
This is consistent with counting from zero-norm states listed in
the table.  Three traceless conditions read from zero-norm states
corresponds to the three equations involving $\d
\xi^{\;\;\;\n}_{\m\n}, \d \xi^{0\;\m}_\m, \d \xi^{1\;\m}_\m$ which
are contained in eqs~(\ref{rule_DDg_a}), (\ref{rule_DDg_c}), and
(\ref{rule_DDg_d}).

It is important to note that the transformation for the matter
parts, eqs~(\ref{rule_CCa})-(\ref{rule_CCd}) and
eqs~(\ref{rule_DDm_a})-(\ref{rule_DDm_f}), are the same as the
calculation \cite{jclee3} based on the chordal gauge
transformation of free covariant string field theory constructed
by Banks and Peskin \cite{Banks}.  The Chordal gauge
transformation can be written in the following form
\be
\d \Phi[X(\s)]= \sum_{n>0} L_{-n} \Phi_n[X(\s)], \label{d_Phi_2}
\ee where $\Phi[X(\s)]$ is the string field and $\Phi_n[X(\s)]$
are gauge parameters which are functions of $X[\s]$ only and free of
ghost fields.  This is because the pure ghost part of $Q_B$ in
eq~(\ref{BRST_C}) does not contribute to the transformation of
matter background fields. It is interesting to note that the
r.h.s. of eq  (23) is in the form of off-shell spurious states \cite{GSW} in the OCFQ
approach. They become zero-norm states on imposing the
physical and on-shell state condition.

Finally, it can be shown that the number of scalar zero-norm
states at n-th massive level ($n \ge 3$) is at least the sum of those at
(n-2)-th  and (n-1)-th massive levels. So positive-norm scalar modes at
n-th level, if they exist, will be decoupled according to our decoupling
conjecture.  The decoupling of these scalars has important implication on
Sen's conjectures on the decay of open string tachyon.  Since all scalars
on D-brane including tachyon get non-zero vev in the false vacuum, they
will decay together with tachyon and disappear eventually to the true
closed string vacuum. As the scalar states together with higher tensor
states form a large gauge multiplet at each mass level, and its scattering amplitudes are fixed
by the tensor fields, these tensor fields of open string (D25-brane) will
accompany the decay process.  This means that the whole D-brane could disappear
to the true closed string vacuum! The mechanism could provide a
hint to solve the so-called $U(1)$-problem \cite{Witten2} in Sen's
conjectures. A further study is in progress.

\section{The fourth massive level} \label{fourth}

We will use both the first and second quantized approaches to test the decoupling
conjecture for the fourth massive level $m^2=8$.
\bde
\item{(A)} The first quantized calculation

The positive-norm physical propagating fields can be found in Ref.~\cite{Manes}.  Their Young tabulations are
\be
\raisebox{0.04in}{\fbox}\hspace{-0.005in}\raisebox{0.04in}{\fbox}\hspace{-0.005in}\raisebox{0.04in}{\fbox}\hspace{-0.005in}\raisebox{0.04in}{\fbox}\hspace{-0.005in}\raisebox{0.04in}{\fbox},\raisebox{0.06in}{\fbox}\hspace{-0.094in}\raisebox{-.029in}{\fbox}\hspace{-0.006in}\raisebox{0.06in}{\fbox}\hspace{-0.006in}\raisebox{0.06in}{\fbox},\raisebox{0.06in}{\fbox}\hspace{-0.094in}\raisebox{-.029in}{\fbox}\hspace{-0.006in}\raisebox{0.06in}{\fbox},\raisebox{0.04in}{\fbox}\hspace{-0.005in}\raisebox{0.04in}{\fbox}\hspace{-0.005in}\raisebox{0.04in}{\fbox},\raisebox{0.06in}{\fbox}\hspace{-0.094in}\raisebox{-.029in}{\fbox},\raisebox{0.04in}{\fbox}.
\label{tableau1} \ee The Young tabulations of zero-norm states can
then be shown to be
\be
\raisebox{0.04in}{\fbox}\hspace{-0.005in}\raisebox{0.04in}{\fbox}\hspace{-0.005in}\raisebox{0.04in}{\fbox}\hspace{-0.005in}\raisebox{0.04in}{\fbox},\raisebox{0.06in}{\fbox}\hspace{-0.094in}\raisebox{-.029in}{\fbox}\hspace{-0.006in}\raisebox{0.06in}{\fbox}^{\;'},2\times\raisebox{0.04in}{\fbox}\hspace{-0.005in}\raisebox{0.04in}{\fbox}\hspace{-0.005in}\raisebox{0.04in}{\fbox},2\times\raisebox{0.06in}{\fbox}\hspace{-0.094in}\raisebox{-.029in}{\fbox},4\times\raisebox{0.04in}{\fbox}\hspace{-0.005in}\raisebox{0.04in}{\fbox},5\times\raisebox{0.04in}{\fbox},3\times\bullet.
\label{tableau2}
\ee

Note that the two representations
$\raisebox{0.06in}{\fbox}\hspace{-0.094in}\raisebox{-.029in}{\fbox}\hspace{-0.006in}\raisebox{0.06in}{\fbox}$
in (\ref{tableau1}) and
$\raisebox{0.06in}{\fbox}\hspace{-0.094in}\raisebox{-.029in}{\fbox}\hspace{-0.006in}\raisebox{0.06in}{\fbox}^{\;'}$
in (\ref{tableau2}) are different.  One corresponds to
$\a^\m_{-1}\a^\n_{-2}\a^\lam_{-2}$ and the other
$\a^\m_{-1}\a^\n_{-1}\a^\lam_{-3}$.  So one expects the last three
states in (\ref{tableau1}) can be
gauged to the higher rank fields.  The most general worldsheet
coupling consistent with vertex operator consideration is
\bea
\hskip -3cm T(\t) = &\;& -\frac{1}{2}\eta_{\m\n}\del_\t X^\m \del_\t X^\n + E_{\m\n\lam\a\b} \del_\t X^\m \del_\t X^\n \del_\t X^\lam \del_\t X^\a \del_\t X^\b +  E_{\m\n\lam\a} \del_\t X^\m \del_\t X^\n \del_\t X^\lam \del_\t^2 X^\a \nonumber \\
\hskip -3cm &\;& + E^0_{\m\n\lam} \del_\t X^\m \del_\t X^\n \del_\t^3 X^\lam + E^1_{\m\n\lam} \del_\t X^\m \del_\t^2 X^\n \del_\t^2 X^\lam + E^0_{\m\n} \del_\t X^\m \del_\t^4 X^\n + E^1_{\m\n} \del_\t^2 X^\m \del_\t^3 X^\n \nonumber \\
&\;& + E_\m \del_\t^5 X^\m. \label{T++5}
\eea
After a lengthy calculation, the condition to cancel all worldsheet q-number
anomalies are
\bse
\bea
&\;& \hspace{-1.1cm} 5\del^\m
E_{\m\n\lam\a\b} - 2E_{(\n\lam\a\b)} = 0, \label{Cond_E1a} \\ &\;&
\hspace{-1.1cm} \del^\m E^0_{\m\n} - 20 E_{\n} = 0, \\ &\;&
\hspace{-1.1cm} \del^\m E_{\m\n\lam\a} - 12E^0_{\n\lam\a} -
8E^1_{(\n\lam)\a}= 0, \\ &\;& \hspace{-1.1cm} \del^\m
E^0_{\m\n\lam} - 6E^0_{\n\lam} - E^1_{\n\lam}= 0, \\ &\;&
\hspace{-1.1cm} \del^\m E^1_{\m\n\lam} - 6E^1_{(\n\lam)} = 0,
\label{Cond_E1e} \eea \ese \vspace{-1.3cm} \bse \bea &\;&
20E^\m_{\;\;\m\n\lam\a} + \del^\m E_{\n\lam\a\m} -
12E^0_{(\n\lam\a)}= 0, \label{Cond_E2a} \\ &\;&
E^{0\m}_{\quad\m\n} + 4\del^\m E^1_{\m\n} - 120E_{\n}= 0, \\ &\;&
E^\m_{\;\;\m\n\lam} + 8\del^\m E^1_{\n\lam\m} - 48 E^0_{\n\lam}-
12E^1_{\lam\n} = 0, \label{Cond_E2c} \eea \ese \vspace{-1.3cm}
\bse \bea &\;& \hspace{-1.4cm} E^\m_{\;\;\n\lam\m} + \del^\m
E^0_{\n\lam\m} - 4E^0_{(\n\lam)}= 0, \label{Cond_E3a} \\ &\;&
\hspace{-1.4cm} E^{1\m}_{\quad\m\n} + 12\del^\m E^1_{\n\m} -
240E_{\n}= 0, \label{Cond_E3b} \eea \ese \vspace{-1.3cm} \bea &\;&
\hspace{-0.3cm} 3E^{0\m}_{\quad\n\m} + E^{1\;\;\m}_{\n\m} +
6\del^\m E^0_{\n\m} - 30E_{\n}= 0, \label{Cond_E4} \eea
\vspace{-1.3cm} \bea &\;& \hspace{-1.8cm} 2E^{0\m}_{\quad\m} +
E^{1\;\;\m}_{\m} + 10\del^\m E_{\m}= 0, \label{Cond_E5} \eea
\vspace{-1.3cm} \bea &\;& \hspace{-4.5cm} (\del^2 - 6) \phi = 0.
\eea
Here, $\phi$ again represents all background fields introduced in eq~(\ref{T++5}). Eqs~(\ref{Cond_E1a})-(\ref{Cond_E1e}) are extracted from $\frac{1}{(\t-\t')^3}$ anomalous terms in the operator product calculation, similarly (\ref{Cond_E2a})-(\ref{Cond_E2c}), (\ref{Cond_E3a})-(\ref{Cond_E3b}), (\ref{Cond_E4}) and (\ref{Cond_E5}) are extracted form $\frac{1}{(\t-\t')^4}, \frac{1}{(\t-\t')^5}, \frac{1}{(\t-\t')^6}$ and $\frac{1}{(\t-\t')^7}$ anomalous terms respectively.  It can be carefully checked, as one did for the third massive level, that only $E_{\m\n\lam\a\b}$ and mixed-symmetric $E_{\m\n\lam\a}$ and $E^1_{\m\n\lam}$ corresponding to the first three Young representations in eq~(\ref{tableau1}) are dynamically independent as the conjecture has claimed. The last three states in eq~(\ref{tableau1}) again can be gauged to the first three states due to the existence of zero-norm states with the same Young representations in eq~(\ref{tableau2}).

\item{(B)} WSFT calculation

$\Phi$ and $\Lam$ can be expanded at this massive level as \bea
&\;& \hspace{-2.5cm} \Phi_5 = \biggl\{ iE_{\m\n\lam\a\b}(x)
\a^\m_{-1}\a^\n_{-1}\a^\a_{-1}a^\lam_{-1}a^\b_{-1} +
E_{\m\n\a\b}(x) \a^\m_{-1}\a^\n_{-1}\a^\a_{-1}a^\b_{-2} -
iE^0_{\m\n\a}(x) \a^\m_{-1}\a^\n_{-1}\a^\a_{-3} \nonumber\\ &\;&
\hspace{-1.3cm} - iE^1_{\m\n\a}(x) \a^\m_{-1}\a^\n_{-2}\a^\a_{-2}
- E^0_{\m\n}(x) \a^\m_{-1} \a^\n_{-4} - E^1_{\m\n}(x) \a^\m_{-2}
\a^\n_{-3} + iE_\m (x) \a^\m_{-5} \nonumber \\ &\;&
\hspace{-1.3cm} + \zeta_{\m\n\a\b}(x) \a^\m_{-1} \a^\n_{-1}
\a^\a_{-1} \a^\b_{-1} b_{-1} c_0 - i\zeta^0_{\m\n\a}(x) \a^\m_{-2}
\a^\n_{-1} \a^\a_{-1} b_{-1} c_0 - i\zeta^1_{\m\n\a}(x) \a^\m_{-1}
\a^\n_{-1} \a^\a_{-1} b_{-2} c_0 \nonumber\\ &\;& \hspace{-1.3cm}
- i\zeta^2_{\m\n\a}(x) \a^\m_{-1} \a^\n_{-1} \a^\a_{-1} b_{-1}
c_{-1} - \zeta^0_{\m\n}(x) \a^\m_{-3} \a^\n_{-1} b_{-1} c_0 -
\zeta^1_{\m\n}(x) \a^\m_{-2} \a^\n_{-2} b_{-1} c_0 \nonumber\\
&\;& \hspace{-1.3cm} - \zeta^2_{\m\n}(x) \a^\m_{-2} \a^\n_{-1}
b_{-2} c_0  - \zeta^3_{\m\n}(x) \a^\m_{-1} \a^\n_{-1} b_{-3} c_0 -
\zeta^4_{\m\n}(x) \a^\m_{-2} \a^\n_{-1} b_{-1} c_{-1} \nonumber\\
&\;& \hspace{-1.3cm} - \zeta^5_{\m\n}(x) \a^\m_{-1} \a^\n_{-1}
b_{-2} c_{-1} - \zeta^6_{\m\n}(x) \a^\m_{-1} \a^\n_{-1} b_{-1}
c_{-2} + i\zeta^0_\m(x) \a^\m_{-4} b_{-1} c_0 + i\zeta^1_\m(x)
\a^\m_{-3} b_{-2} c_0 \nonumber \\ &\;& \hspace{-1.3cm} +
i\zeta^2_\m(x) \a^\m_{-2} b_{-3} c_0 + i\zeta^3_\m(x) \a^\m_{-1}
b_{-4} c_0 + i\zeta^4_\m(x) \a^\m_{-3} b_{-1} c_{-1} +
i\zeta^5_\m(x) \a^\m_{-2} b_{-2} c_{-1} \nonumber\\ &\;&
\hspace{-1.3cm} + i\zeta^6_\m(x) \a^\m_{-1} b_{-3} c_{-1} +
i\zeta^7_\m(x) \a^\m_{-2} b_{-1} c_{-2} + i\zeta^8_\m(x)
\a^\m_{-1} b_{-2} c_{-2} + i\zeta^9_\m(x) \a^\m_{-1} b_{-1} c_{-3}
\nonumber\\ &\;& \hspace{-1.3cm} + i\zeta^{10}_\m(x) \a^\m_{-1}
b_{-2} b_{-1} c_{-1} c_{0} + \zeta^0(x) b_{-5} c_0 + \zeta^1(x)
b_{-4} c_{-1} + \zeta^2(x) b_{-3} c_{-2} + \zeta^3(x) b_{-2}
c_{-3} \nonumber \\ &\;& \hspace{-1.3cm} + \zeta^4(x) b_{-1}
c_{-4} + \zeta^5(x) b_{-3} b_{-1} c_{-1} c_{-0} + \zeta^6(x)
b_{-2} b_{-1} c_{-2} c_{-0} \biggr\}c_1 |k\ra \\ &\;&
\hspace{-2.5cm} \Lam_5 = \biggl\{ \eps^0_{\m\n\a\b}(x) \a^\m_{-1}
\a^\n_{-1} \a^\a_{-1} \a^\b_{-1} b_{-1} - i\eps^1_{\m\n\a}(x)
\a^\m_{-2} \a^\n_{-1} \a^\a_{-1} b_{-1} - i\eps^2_{\m\n\a}(x)
\a^\m_{-1} \a^\n_{-1} \a^\a_{-1} b_{-2} \nonumber\\ &\;&
\hspace{-1.3cm} - \eps^3_{\m\n}(x) \a^\m_{-3} \a^\n_{-1} b_{-1} -
\eps^4_{\m\n}(x) \a^\m_{-2} \a^\n_{-2} b_{-1} - \eps^5_{\m\n}(x)
\a^\m_{-2} \a^\n_{-1} b_{-2} - \eps^6_{\m\n}(x) \a^\m_{-1}
\a^\n_{-1} b_{-3} \nonumber\\ &\;& \hspace{-1.3cm} -
\eps^7_{\m\n}(x) \a^\m_{-1} \a^\n_{-1} b_{-2} b_{-1} c_{0} +
i\eps^7_{\m}(x) \a^\m_{-4} b_{-1} + i\eps^8_{\m}(x) \a^\m_{-3}
b_{-2} + i\eps^9_{\m}(x) \a^\m_{-2} b_{-3} + i\eps^{10}_{\m}(x)
\a^\m_{-1} b_{-4} \nonumber\\ &\;& \hspace{-1.3cm} +
i\eps^{11}_{\m}(x) \a^\m_{-2} b_{-2} b_{-1} c_0 +
i\eps^{12}_{\m}(x) \a^\m_{-1} b_{-3} b_{-1} c_0 +
i\eps^{13}_{\m}(x) \a^\m_{-1} b_{-2} b_{-1} c_{-1} + \eps^7(x)
b_{-5} \nonumber\\ &\;& \hspace{-1.3cm} + \eps^8(x) b_{-4} b_{-1}
c_0 + \eps^9(x) b_{-3} b_{-2} c_0 + \eps^{10}(x) b_{-3}b_{-1}
c_{-1} + \eps^{11}(x) b_{-2}b_{-1} c_{-2} \biggr\} |\Omega\ra.
\eea The transformations for the matter part are \bse\bea &\;&
\hspace{-4.5cm} \d E_{\m\n\lam\a\b} = -  \del_{(\b}
\eps^0_{\m\n\lam\a)} + \frac{1}{2} \eps^2_{(\lam\a\b}
\eta_{\m\n)}, \\ &\;& \hspace{-4.5cm} \d E_{\m\n\a\b} = -
\del_{(\m} \eps^1_{|\b|\a\n)} -  \del_{\b} \eps^2_{\a\m\n} +
4\eps^0_{\m\n\a\b} - \frac{1}{2} \eps^5_{\b(\n} \eta_{\a\m)} -
\eps^6_{(\a\m} \eta_{\n)\b}, \\ &\;& \hspace{-4.5cm} \d
E^0_{\m\n\a} = - \del_{(\m} \eps^3_{|\a|\n)} - \del_{\a}
\eps^6_{\n\m} + 2 \eps^1_{\a\n\m} + 3 \eps^2_{\a\n\m} -
\frac{1}{2} \eps^7_{\a} \eta_{\n\m} - \eps^9_{(\m} \eta_{\n)\a},\\
&\;& \hspace{-4.5cm} \d E^1_{\m\n\a} = - \del_{\m} \eps^4_{\n\a} -
\del_{(\a} \eps^5_{\n)\m} + 2 \eps^1_{(\a\n)\m} - \eps^8_{(\a}
\eta_{\n)\m} - \frac{1}{2} \eps^9_{\m} \eta_{\n\a},\\ &\;&
\hspace{-4.5cm} \d E^0_{[\m\n]} = - \del_{[\m} \eps^7_{\n]} -
\del_{[\n} \eps^9_{\m]} + 3\eps^3_{[\n\m]} + 2\eps^5_{[\n\m]},\\
&\;& \hspace{-4.5cm} \d E^1_{[\m\n]} = - \del_{[\m} \eps^7_{\n]} -
\del_{[\n} \eps^8_{\m]} + \eps^3_{[\n\m]} + \eps^5_{[\m\n]},\\
&\;& \hspace{-4.5cm} \d E^0_{(\m\n)} = - \del_{(\m} \eps^7_{\n)} -
\del_{(\n} \eps^9_{\m)} + 3\eps^3_{(\n\m)} + 2\eps^5_{(\n\m)} +
2\eps^6_{\n\m} - \eps^7 \eta_{\m\n},\\ &\;& \hspace{-4.5cm} \d
E^1_{(\m\n)} = - \del_{(\m} \eps^7_{\n)} -  \del_{(\n}
\eps^8_{\m)} + \eps^3_{(\n\m)} + 2\eps^4_{\n\m} + \eps^5_{(\m\n)}
- \eps^7 \eta_{\m\n},\\ &\;& \hspace{-4.5cm} \d E_{\m} = - \del_\m
\eps^7 + 7\eps^7_\m + 2\eps^8_\m + \eps^9_{\m}. \eea\ese \ede
Again these are the same as the calculation by eq~(\ref{d_Phi_2}).
All background fields except $E_{\m\n\lam\a\b}$,
mixed-symmetric $E_{\m\n\lam\a}$ and $E^1_{\m\n\lam}$ can be
either gauged away or gauged to $E_{\m\n\lam\a\b}$,
$E_{\m\n\lam\a}$ and $E^1_{\m\n\lam}$ by zero-norm states. This
is consistent with the result of the first quantized approach
presented in subsection (A).  The transformation for the ghost
part is very lengthy and is given in the appendix. There are 18
on-mass-shell conditions, which contains a spin four, a
mixed-symmetric spin three, two symmetric spin three, two
antisymmetric spin two, four symmetric spin two, five vector and
three scalar fields, and 15 gauge conditions. It is again consistent
with counting the number of zero-norm states listed in
eq~(\ref{tableau2}).

\section{Conclusion}

We have explicitly shown that the degenerate positive-norm states
at the third and fourth massive levels of bosonic open string
theory can be gauged to the higher rank fields at the same mass
level. This means that the scattering amplitudes of these
degenerate positive-norm states can be expressed in terms of those
of higher spin states at the same mass level through massive Ward
identities. This is demonstrated by using both OCFQ string and
WSFT.  We have compared the on-shell conditions of zero-norm
states in OCFQ stringy gauge transformation to the background
ghost fields in off-shell gauge transformation of WSFT. This
important observation makes the lengthy calculations in both the
first and second quantized approaches controllable and more
importantly provides a double consistency check of our results. The
interesting stringy behaviors discussed in this paper and those in
Ref.~\cite{Gross1,Gross2} seem to imply that there must exist
enormous exotic high-energy properties of string theory which
remained to be uncovered.  One interesting application of the
decoupling of higher scalar modes is the decay of tensor fields on
D-brane into the true closed string vacuum in Sen's conjectures.

It is straightforward to generalize our calculation to closed string theory for the first quantized approach presented in sections II and IV (A). Another way to generalize to the closed string case is to make use of the simple relation between closed and open string amplitudes in [20].  A reliable second quantized closed string field theory may help uncover more high energy stringy properties.

\section*{Acknowledgment}

The authors would like to thank Pei-ming Ho for stimulating discussions which help clarify some of the points in the paper. The authors also thank Miao Li for helpful conversations. This work is supported in part by the National Science Council, the Center for Theoretical Physics at National Taiwan University, the National Center for Theoretical Sciences, and the CosPA project of the Ministry of Education, Taiwan, R.O.C.

\vfill\eject
$\quad$
\vspace{-2cm}
\section*{Appendix}
Gauge transformation for background ghost fields of the fourth massive level are:
\setcounter{equation}{0}
\renewcommand{\theequation}{A.\arabic{equation}}
\bea
&\;& \d \zeta_{\m\n\a\b} = \frac{1}{2}( \del^2 - 8)\eps^0_{\m\n\a\b} - \frac{1}{2} \eps^7_{(\m\n} \eta_{\a\b)}, \label{rule_EEg1} \\
&\;& \d \zeta^0_{\m\n\a} = \frac{1}{2}( \del^2 - 8)\eps^1_{\m\n\a} -  \del_{\m} \eps^7_{\n\a} - \frac{1}{2} \eps^{11}_{\m} \eta_{\n\a} - \eps^{12}_{(\n} \eta_{\a)\m}, \label{rule_EEg2} \\
&\;& \d \zeta^1_{\m\n\a} = \frac{1}{2}( \del^2 - 8)\eps^2_{\m\n\a} +  \del_{(\m} \eps^7_{\n\a)},\\
&\;& \d \zeta^2_{\m\n\a} = -4 \del^{\b} \eps^0_{\m\n\a\b} - 2\eps^1_{(\m\n\a)} - 3\eps^2_{\m\n\a} - \frac{1}{2} \eps^{13}_{(\m} \eta_{\n\a)}, \\
&\;& \d \zeta^0_{[\m\n]} = \frac{1}{2}( \del^2 - 8)\eps^3_{[\m\n]} -  \del_{[\m} \eps^{12}_{\n]}, \\
&\;& \d \zeta^2_{[\m\n]} = \frac{1}{2}( \del^2 - 8)\eps^5_{[\m\n]} +  \del_{[\n} \eps^{11}_{\m]}, \\
&\;& \d \zeta^4_{[\m\n]} = -2 \del^{\a} \eps^1_{[\m\n]\a} -  \del_{[\m} \eps^{13}_{\n]} - 3\eps^3_{[\m\n]} - 3\eps^5_{[\m\n]}, \\
&\;& \d \zeta^0_{(\m\n)} = \frac{1}{2}( \del^2 - 8)\eps^3_{(\m\n)} -  \del_{(\m} \eps^{12}_{\n)} + 2\eps^7_{\m\n} - \eps^8 \eta_{\m\n}, \\
&\;& \d \zeta^1_{\m\n} = \frac{1}{2}( \del^2 - 8)\eps^4_{\m\n} -  \del_{(\m} \eps^{11}_{\n)} - \frac{1}{2}\eps^8 \eta_{\a\b}, \\
&\;& \d \zeta^2_{(\m\n)} = \frac{1}{2}( \del^2 - 8)\eps^5_{(\m\n)} +  \del_{(\n} \eps^{11}_{\m)} - 2\eps^7_{\m\n} - \eps^9 \eta_{\m\n}, \\
&\;& \d \zeta^3_{\m\n} = \frac{1}{2}( \del^2 - 8)\eps^6_{\m\n} +  \del_{(\m} \eps^{12}_{\n)} - \eps^7_{\m\n} + \frac{1}{2}\eps^9 \eta_{\m\n}, \\
&\;& \d \zeta^4_{(\m\n)} = -2 \del^{\a} \eps^1_{(\m\n)\a} -  \del_{(\m} \eps^{13}_{\n)} - 3\eps^3_{(\m\n)} - 4\eps^4_{\m\n} - 3\eps^5_{(\m\n)} - \eps^{10} \eta_{\m\n}, \\
&\;& \d \zeta^5_{\m\n} = -3 \del^{\a} \eps^2_{\m\n\a} +  \del_{(\m} \eps^{13}_{\n)}- 2\eps^5_{(\m\n)} - 4\eps^6_{\m\n} - 2\eps^7_{\m\n}, \\
&\;& \d \zeta^6_{\m\n} = -2 \del^{\a} \eps^1_{\a\m\n} + 6\eps^0_{\m\n\a\b}\eta^{\a\b} - 3\eps^3_{(\m\n)} - 5\eps^6_{\m\n} + 4\eps^7_{\m\n} - \frac{1}{2} \eps^{11} \eta_{\m\n}, \\
&\;& \d \zeta^0_\m = \frac{1}{2}( \del^2 - 8)\eps^7_{\m} -  \del_{\m} \eps^8 + 2\eps^{11}_{\m} + \eps^{12}_{\m}, \\
&\;& \d \zeta^1_\m = \frac{1}{2}( \del^2 - 8)\eps^8_{\m} -  \del_{\m} \eps^9 - 2\eps^{11}_{\m}, \\
&\;& \d \zeta^2_\m = \frac{1}{2}( \del^2 - 8)\eps^9_{\m} +  \del_{\m} \eps^9 - \eps^{12}_{\m}, \\
&\;& \d \zeta^3_\m = \frac{1}{2}( \del^2 - 8)\eps^{10}_{\m}+  \del_{\m} \eps^8 - 2\eps^{12}_{\m}, \\
&\;& \d \zeta^4_\m = - \del^\n \eps^3_{\m\n} -  \del_\m \eps^{10} - 4\eps^7_\m - 3\eps^8_\m + \eps^{13}_{\m}, \\
&\;& \d \zeta^5_\m = - \del^\n \eps^5_{\m\n} - 3\eps^8_\m - 4\eps^9_{\m} - 2\eps^{11}_{\m} - \eps^{13}_{\m}, \\
&\;& \d \zeta^6_\m = -2 \del^\n \eps^6_{\m\n} +  \del_\m \eps^{10}- 2\eps^9_{\m} - 5\eps^{10}_{\m} - 2\eps^{12}_{\m}- \eps^{13}_\m ,
\eea
\bea
&\;& \d \zeta^7_\m = -4 \del^\n \eps^4_{\m\n} -  \del_\m \eps^{11}- 4\eps^7_{\m} - 5\eps^{9}_{\m} + 4\eps^{11}_{\m} + \eps^{1}_{\m\n\a} \eta^{\n\a}, \\
&\;& \d \zeta^8_\m = -2 \del^\n \eps^5_{\n\m} +  \del_\m \eps^{11}- 3\eps^8_{\m} - 6\eps^{10}_{\m} - 3\eps^{13}_{\m} + 3\eps^{2}_{\m\n\a} \eta^{\n\a}, \\
&\;& \d \zeta^9_\m = -3 \del^\n \eps^3_{\m\n} - 4\eps^7_{\m} - 7\eps^{10}_{\m} + 6\eps^{12}_{\m} + 5\eps^{13}_\m + 4\eps^{1}_{\n\a\m} \eta^{\n\a}, \\
&\;& \d \zeta^{10}_\m = \frac{1}{2}( \del^2 - 8)\eps^{13}_{\m} + 2 \del^\n \eps^7_{\m\n} + 2\eps^{11}_{\m} + 4\eps^{12}_{\m}, \\
&\;& \d \zeta^0 = \frac{1}{2}( \del^2 - 8)\eps^7 - 3\eps^8 - \eps^9, \\
&\;& \d \zeta^1 = - \del^\m \eps^{10}_{\m} - 6\eps^7 - 2\eps^8 - 2\eps^{10}, \\
&\;& \d \zeta^2 = - \del^\m \eps^9_{\m} - 7\eps^7 - 4\eps^9 - 3\eps^{10} - \eps^{11} + \eps^6_{\m\n} \eta^{\m\n}, \\
&\;& \d \zeta^3 = -3 \del^\m \eps^8_{\m} - 8\eps^7 + 6\eps^9 - 4\eps^{11} + 2\eps^5_{\m\n} \eta^{\m\n}, \\
&\;& \d \zeta^4 = -4 \del^\m \eps^7_{\m} - 9\eps^7 + 8\eps^8 + 7\eps^{10} + 6\eps^{11} + 3\eps^3_{\m\n} \eta^{\m\n} + 4\eps^4_{\m\n} \eta^{\m\n}, \\
&\;& \d \zeta^5 = \frac{1}{2}( \del^2 - 8)\eps^{10} +  \del^\m \eps^{12}_{\m} + 5\eps^8 + 3\eps^9, \\
&\;& \d \zeta^6 = \frac{1}{2}( \del^2 - 8)\eps^{11} + 2 \del^\m \eps^{11}_{\m} + 6\eps^8 - 5\eps^9 - \eps^7_{\m\n} \eta^{\m\n}.  \label{rule_EEg32}
\eea
There are 18 on-mass-shell conditions and 15 gauge conditions in eqs~(\ref{rule_EEg1})-(\ref{rule_EEg32}), which are consistent with counting from number of zero-norm states listed in eq~(\ref{tableau2}). Note that there are two irreducible components in (\ref{rule_EEg2}).

\end{document}